\documentclass[english,reprint,amsmath,amssymb,aps,setspace,pre,superscriptaddress]{revtex4-1}
\usepackage[latin9]{inputenc}
\usepackage{babel}
\usepackage{bm}
\usepackage{amsmath}
\usepackage{units}
\usepackage{amstext}
\usepackage{amssymb}

\usepackage{graphicx}
\usepackage[unicode=true]
 {hyperref}

\makeatletter
\usepackage{graphicx}
\usepackage{dcolumn}
\usepackage{bm}
\usepackage{setspace}
\usepackage{breqn}
\usepackage{hyperref}

\let\cat@comma@active\@empty

\gdef\wrap@breqn@environ#1#2{
\expandafter\let\csname breqn@oldbegin@#1\expandafter\endcsname\csname #1\endcsname
\expandafter\let\csname breqn@oldend@#1\expandafter\endcsname\csname end#1\endcsname
\expandafter\gdef\csname breqn@begin@#1\endcsname{%
\expandafter\let\csname #1\expandafter\endcsname\csname breqn@oldbegin@#1\endcsname%
\begin{#2}%
}
\expandafter\gdef\csname breqn@end@#1\endcsname{%
\expandafter\let\csname end#1\expandafter\endcsname\csname breqn@oldend@#1\endcsname%
\end{#2}%
\expandafter\let\csname #1\expandafter\endcsname\csname breqn@begin@#1\endcsname%
\expandafter\let\csname end#1\expandafter\endcsname\csname breqn@end@#1\endcsname%
}
\expandafter\let\csname #1\expandafter\endcsname\csname breqn@begin@#1\endcsname
\expandafter\let\csname end#1\expandafter\endcsname\csname breqn@end@#1\endcsname
}
\wrap@breqn@environ{equation}{dmath}
\wrap@breqn@environ{equation*}{dmath*}

\newcounter{examplecounter}
\newenvironment{example}{%
    \refstepcounter{examplecounter}%
}

\makeatother

\begin{document}

\title{Modularity and the spread of perturbations in complex dynamical systems}

\author{Artemy Kolchinsky}

\email{akolchin@indiana.edu}

\affiliation{School of Informatics and Computing, Indiana University, Bloomington,
Indiana 47408, USA}

\affiliation{Program in Cognitive Science, Indiana University, Bloomington, Indiana
47408, USA}

\author{Alexander J. Gates}

\affiliation{School of Informatics and Computing, Indiana University, Bloomington,
Indiana 47408, USA}

\affiliation{Program in Cognitive Science, Indiana University, Bloomington, Indiana
47408, USA}

\author{Luis M. Rocha}

\affiliation{School of Informatics and Computing, Indiana University, Bloomington,
Indiana 47408, USA}

\affiliation{Program in Cognitive Science, Indiana University, Bloomington, Indiana
47408, USA}

\affiliation{Instituto Gulbenkian de Ciência, Rua da Quinta Grande, 6, 2780-156
Oeiras, Portugal}
\begin{abstract}
We propose a method to decompose dynamical systems based on the idea
that modules constrain the spread of perturbations. We find partitions
of system variables that maximize `perturbation modularity', defined
as the autocovariance of coarse-grained perturbed trajectories. The
measure effectively separates the fast intramodular from the slow
intermodular dynamics of perturbation spreading (in this respect,
it is a generalization of the `Markov stability' method of network
community detection). Our approach captures variation of modular organization
across different system states, time scales, and in response to different
kinds of perturbations: aspects of modularity which are all relevant
to real-world dynamical systems. It offers a principled alternative
to detecting communities in networks of statistical dependencies between
system variables (e.g., `relevance networks' or `functional networks').
Using coupled logistic maps, we demonstrate that the method uncovers
hierarchical modular organization planted in a system's coupling matrix.
Additionally, in homogeneously-coupled map lattices, it identifies
the presence of self-organized modularity that depends on the initial
state, dynamical parameters, and type of perturbations. Our approach
offers a powerful tool for exploring the modular organization of complex
dynamical systems.
\end{abstract}
\maketitle
Many complex systems are \emph{modular}, in that their components
are organized in tightly-integrated subsystems that are weakly coupled
to one another. Modularity has been argued to play many important
roles, including increasing robustness \cite{simon1962thearchitecture,kitano2004biological,wu2009response},
evolvability \cite{simon1962thearchitecture,wagner1996complex}, and
functional differentiation \cite{tononi1994ameasure,wagner2007theroad}.
Thus, there is great interest in measures of modularity and methods
for decomposing complex systems into weakly coupled modules.

This problem is here considered in the domain of multivariate dynamics,
a common formalism for modeling complex physical, biological, neural,
and social systems. We propose a method of identifying dynamical modules
motivated by the intuition that, in a modular system, the spread of
perturbations is characterized by two time scales: fast spreading
within modules and slow spreading between modules \cite{simon1962thearchitecture,pan2009modularity}.
In our treatment, the spreading process is coarse-grained relative
to a \emph{partition} (a decomposition of system variables into disjoint
subsystems) by measuring the magnitude of the perturbation's effect
within each subsystem over time. If a partition reflects underlying
modular structure, initially perturbed subsystems remain affected
as dynamics unfold, while initially unperturbed subsystems remain
largely unaffected. In this case, the partition's coarse graining
will capture the slow component of perturbation spreading dynamics,
an effect quantified using a quality function called \emph{perturbation
modularity}. Our perturbation-based approach is related to many existing
techniques for analyzing multivariate dynamics, including Lyapunov-exponent
based methods \cite{luque2000lyapunov,pomerance2009theeffect,strogatz2014nonlinear}
and impulse response analysis \cite{koop1996impulse}.

As will be elaborated below, our methodology can identify the dependence
of optimal decompositions on initial states, time scales, and kinds
of perturbations applied. These factors are all important aspects
of modular organization in real-world dynamical systems. Dependence
on the initial condition reflects that dynamical systems can exhibit
different modular organizations in different regions of their state-space;
for example, distributed regions of the brain can couple into modular
assemblies via oscillatory synchronization, with the same region participating
in different assemblies depending on brain state \cite{fries2005amechanism,womelsdorf2007modulation}.
The choice of time scale affects optimal decompositions by determining
the separation between intramodular and intermodular perturbation
spreading; in real-world complex systems, longer time scales have
often been argued to correspond to larger-scale modules \cite{simon1962thearchitecture,wu1995frombalance,arenas2006sync,palla2007quantifying,bassett2011dynamic}.
Finally, the dependence on the kinds of perturbations reflects that
a dynamical system may be robust to some perturbations but highly-sensitive
to others \cite{carlson2002complexity}; for example, in biological
double-knockout experiments, cellular responses to the simultaneous
deactivation of two genes can differ dramatically from responses to
the individual deactivation of either gene \cite{deutscher2008geneknockout}.

Our approach starts from a pre-specified dynamical system and thus
differs fundamentally from existing treatments of modularity based
on network representations of a system. Such methods are usually unable
to capture the variation of modular organization across state-space
or time scale, as well as other important dynamical aspects of modularity.

For instance, one standard approach applies graph-based\emph{ community
detection} \cite{fortunato2010community} to the structural network
underlying a dynamical process (e.g., the social network over which
an epidemic spreads). This treatment ignores the fact that the same
structural network can support many different dynamical processes
(for example, `complex contagion' epidemics proceed differently from
`simple contagion' epidemics \cite{centola2007complex}). In contrast,
our methodology is by definition sensitive to dynamical differences. 

Another class of methods detects community structure in network representations
of dynamics, defined either in terms of causal interactions or statistical
dependencies between variables (e.g., \emph{relevance networks }in
systems biology \cite{zhang2005ageneral} and \emph{functional networks
}in neuroscience \cite{rubinov2010complex}). Unfortunately, constructing
such networks involves a conversion of the dynamical system (defined
in terms of transitions between multidimensional states) into a graph
(defined in terms of nodes and edges). This conversion can affect
modular decompositions in opaque ways as well as invalidate presumed
graph-theoretic null models \cite{zalesky2012onthe,MacMahon2015CommCorr};
statistical dependency networks in particular require a number of
nontrivial decisions regarding the choice of dependency measure (correlation,
transfer entropy, phase-locking measures, etc.), treatment of positive
versus negative interactions, and thresholding \cite{rubinov2010complex}.
Furthermore, coupling between variables does not necessarily give
rise to large values of correlation or other dependency measures \cite{soriano2012synchronization}
(also as shown in our first example below). Finally, community detection
on dependency networks optimizes quality functions that are difficult
to interpret in terms of the original system dynamics. Perturbation
modularity does not require the construction of a network representation
of a dynamical system and is interpretable in terms of the separation
of slow and fast time scales of perturbation spreading.

Because our approach is based on intrinsic system dynamics, it also
differs from methods that identify modules by \emph{imposing }a dynamical
process onto a given network, such as diffusion of random walkers
\cite{rosvall2008mapsof,delvenne2010stability} or coupled phase oscillators
\cite{arenas2006sync,boccaletti2007detecting}. However, as we discuss
below, in certain cases our approach has connections to such methods.
In particular, it can be seen as a generalization of the random-walk-based
approach of \emph{Markov stability }\cite{delvenne2010stability,lambiotte2014random}
to a broad class of dynamics.

To formally define perturbation modularity, consider a dynamical system
with an $N$-dimensional state-space $\bm{X}$ and evolution operator
$f^{t}:\bm{X}\rightarrow\bm{X}$ at time scale $t$ (both state and
time can be continuous or discrete). Given a set $\mathcal{E}$ of
possible initial perturbations, $\bm{\varepsilon}\in\mathcal{E}$
is applied to an initial condition $\bm{x}\in\bm{X}$ to produce a
perturbed initial condition $\bm{x}+\bm{\varepsilon}\in\bm{X}$. After
time $t$, the size of the perturbation in the whole system is measured
as the norm of the difference between the perturbed and unperturbed
trajectories: $\left\Vert f^{t}(\bm{x}+\bm{\varepsilon})-f^{t}(\bm{x})\right\Vert $.
The relative size of the perturbation within a \emph{subsystem} $S$
(a subset of system variables) is: 
\begin{equation}
m_{S}^{t}(\bm{x},\bm{\varepsilon})=\frac{\left\Vert \left(f^{t}(\bm{x}+\bm{\varepsilon})-f^{t}(\bm{x})\right)_{S}\right\Vert }{\left\Vert f^{t}(\bm{x}+\bm{\varepsilon})-f^{t}(\bm{x})\right\Vert }\label{eq:pertmod-m}
\end{equation}
where the subscript $S$ on the right hand side indicates a projection
onto the dimensions indexed by $S$. For simplicity, we consider only
cases where the system's perturbed and unperturbed trajectories have
not merged ($\left\Vert f^{t}(\bm{x}+\bm{\varepsilon})-f^{t}(\bm{x})\right\Vert >0$)
and Eq.~\eqref{eq:pertmod-m} is well-defined.

Assume a partition of the system $\pi=\left\{ S_{1},\dots,S_{K}\right\} $
into $K$ disjoint subsystems. The \emph{coarse-grained perturbation
vector} $\bm{y}_{\pi}^{t}(\bm{x},\bm{\varepsilon})=\left\langle m_{S_{1}}^{t}(\bm{x},\bm{\varepsilon}),\dots,m_{S_{K}}^{t}(\bm{x},\bm{\varepsilon})\right\rangle $
captures the relative size of the perturbation in each subsystem.
Due to the normalization in Eq.~\eqref{eq:pertmod-m}, $\bm{y}_{\pi}^{t}$
is invariant to the dynamical expansion of the whole-system phase-space,
instead reflecting only the relative effects of perturbations on different
subsystems.

We now define perturbation modularity $Q^{t}(\pi,\bm{x})$ as the
vector autocovariance of the coarse-grained perturbation vector:
\begin{flalign}
 & Q^{t}(\pi,\bm{x})=\label{eq:Q}\\
 & \;\;\;\;\textrm{E}\!\left[\bm{y}_{\pi}^{0}(\bm{x},\bm{\varepsilon})\cdot\bm{y}_{\pi}^{t}(\bm{x},\bm{\varepsilon})\right]-\mathbb{\textrm{E}}\!\left[\bm{y}_{\pi}^{0}(\bm{x},\bm{\varepsilon})\right]\cdot\textrm{E}\!\left[\bm{y}_{\pi}^{t}(\bm{x},\bm{\varepsilon})\right]\nonumber 
\end{flalign}
where the expected values are taken over $P(\bm{\varepsilon})$, a
probability distribution over perturbations (i.e. the elements of
$\mathcal{E}$). The first term of Eq.~\eqref{eq:Q} measures the
degree to which perturbations persist within a partition's subsystems
(i.e.\ initially perturbed subsystems remain affected after time
$t$, while initially unperturbed subsystems remain relatively unaffected).
The second term of Eq.~\eqref{eq:Q} provides a baseline expectation
of perturbation effects that accounts for differences in subsystem
sizes.

As stated, the spread of perturbations in a modular system will be
constrained by module boundaries. The optimal modular decomposition
is the partition that maximizes perturbation modularity: ${\pi^{\star}=\arg\max_{\pi}Q^{t}(\pi,\bm{x})}$.

Perturbation modularity (Eq.~\eqref{eq:Q}), as well as optimal modular
decompositions, are state-dependent in that they are defined relative
to an initial condition $\bm{x}.$ Different criteria may be used
to determine the choice of this initial condition, such as dynamical
importance (e.g., an equilibrium state), particular research interest,
or random selection. Alternatively, the modularity of entire state-space
regions, rather than individual states, can be measured as the expectation
of perturbation modularity over a distribution of initial conditions
(e.g., by averaging across the entire system state space). Similarly,
stochastic dynamical systems can be accommodated by taking expectations
over future state distributions. For simplicity, however, these extensions
are not considered in the present work.

In addition to initial condition, perturbation modularity and optimal
decompositions also depend on the time scale $t$, which, as mentioned,
can act as a resolution parameter. When there is not a time scale
of\emph{ a priori} interest, optimal decompositions can be identified
at multiple resolutions by sweeping across a range of time scales.
Finally, the measure also depends on the kinds of perturbations applied,
as specified by $\mathcal{E}$ and $P(\bm{\varepsilon})$. In practice,
perturbations can be selected according to domain knowledge (e.g.,
typically encountered environmental perturbations) or using `neutral'
options (e.g., small increments to single variables). In many cases,
initial perturbations should be localized to a small number of variables
(i.e. the elements of $\mathcal{E}$ are sparse) because the spread
of perturbations is more pronounced when only a few subsystems are
initially perturbed. As we will show, perturbations that simultaneously
affect many variables probe the system at larger scales and uncover
larger modules, providing another way to explore decompositions at
multiple resolutions.

Like other temporally-localized methods \cite{pikovsky1998dynamic},
perturbation modularity also depends on the norm used to measure perturbation
magnitude. Below, the $\ell_{1}$ norm is used because it performs
well and permits connections to community detection methods in graphs
(see Supplemental Material \footnote{See Supplemental Material for derivation of bounds on perturbation
modularity, connection to Markov stability community detection, and
mapping to Newman's modularity.} for definition of $\ell_{p}$ norms).

Perturbation modularity is related to the \emph{Markov stability }method
of community detection in graphs, which identifies communities as
subgraphs that trap random walkers \cite{delvenne2010stability,lambiotte2014random}.
Similarly to perturbation modularity, Markov stability separates diffusion
dynamics into fast intracommunity and slow intercommunity components.
As shown in the Supplemental Material~\cite{Note1}, perturbation
modularity is equivalent to Markov stability when the system of interest
exhibits diffusion dynamics, perturbations are homogenous increases
to single variables, and the $\ell_{1}$ norm is used to measure perturbation
magnitude. More broadly, our approach can be seen as a generalization
of Markov Stability to other dynamics.

In addition, $\ell_{1}$ perturbation modularity on a dynamical system
is equivalent to \emph{directed weighted Newman's modularity} \cite{arenas2007sizereduction,leicht2008community}
on a specially constructed graph (see Supplemental Material~\cite{Note1}).
In this graph, nodes correspond to system variables and the edge from
node $i$ to node $j$ has weight:
\[
w_{ij}=\textrm{E}\!\left[m_{\{i\}}^{0}(\bm{x},\bm{\varepsilon})\thinspace m_{\{j\}}^{t}(\bm{x},\bm{\varepsilon})\right]
\]
where the expectation is over $P(\bm{\varepsilon})$ and the subscripts
$\{i\}$ and $\{j\}$ indicate single-variable subsystems. This mapping
permits perturbation modularity to be maximized using highly efficient
existing community detection algorithms (such as the Louvain method
\cite{blondel2008fastunfolding,aldecoa2014surpriseme} used for the
examples below; code is available online \footnote{Example code for identifying optimal decompositions at \texttt{\href{https://github.com/artemyk/perturbationmodularity/}{https://github.com/artemyk/perturbationmodularity/}}}).

Several criteria can be used to measure the quality of identified
decompositions. High-quality decompositions have large perturbation
modularity values (e.g., near 1 for $\ell_{1}$ or $\ell_{2}$ perturbation
modularity, see Supplemental Material~\cite{Note1} for derivation
of bounds on perturbation modularity). Additionally, high-quality
decompositions are robust to small changes in system and optimization
parameters. This can be quantified by measures of partition similarity
like normalized mutual information (NMI) \cite{danon2005comparing},
an information-theoretic measure that ranges from 0 (maximally dissimilar
partitions) to 1 (identical partitions). In several of the examples
below, we plot NMI similarity between optimal decompositions identified
at close values of $t$; high NMI values indicate modular organization
robust to small changes in time scale. Similar techniques are used
in the Markov stability literature to identify time scales with robust
decompositions \cite{schaub2012markovdynamics}.

We demonstrate our method on several examples of \emph{coupled logistic
maps}, nonlinear discrete-time dynamical systems that have been used
to explore spatially-extended chaos and pattern formation \cite{kaneko1989pattern}.
Assume a system of $N$ variables, with $x_{i}(t)$ indicating the
state of variable $i$ at time $t$, and the transition function:
\begin{equation}
x_{i}(t+1)=(1-\gamma)g(x_{i}(t))+\gamma\sum_{j\ne i}\frac{k_{ji}}{d_{i}}g(x_{j}(t))\label{eq:cml-equation}
\end{equation}
where $g(x)=1-\alpha x^{2}$ is the logistic map, parameter $\alpha\in[1,2]$
controls the chaoticity, parameter $\gamma\in[0,1]$ controls the
coupling strength, `coupling matrix' elements $k_{ji}$ determine
the influence of variable $j$ on variable $i$, and $d_{i}=\sum_{j\ne i}k_{ji}$
normalizes the coupling strengths. When variables are homogeneously
coupled to nearest neighbors on a 1-dimensional ring lattice, these
systems are called \emph{coupled map lattices} (CML) \cite{kaneko1989pattern}.
Coupled logistic maps display a rich variety of spatiotemporal patterns
in different parameter regimes due to the interplay between intervariable
coupling (which `homogenizes' variable states) and chaos (which injects
variation into variable states).

We consider several examples of coupled logistic maps. Unless otherwise
noted, perturbations consist of a uniform distribution over small
increases to single variables: $\mathcal{E}=\left\{ 0.0001\cdot{\bf e}_{i}:i=1..N\right\} $,
where $\mathbf{e}_{i}$ is the $i^{\textrm{th}}$ $N$-dimensional
standard basis vector. The $\ell_{1}$ norm is used to measure perturbation
size.

\begin{figure}
\includegraphics[width=1\columnwidth]{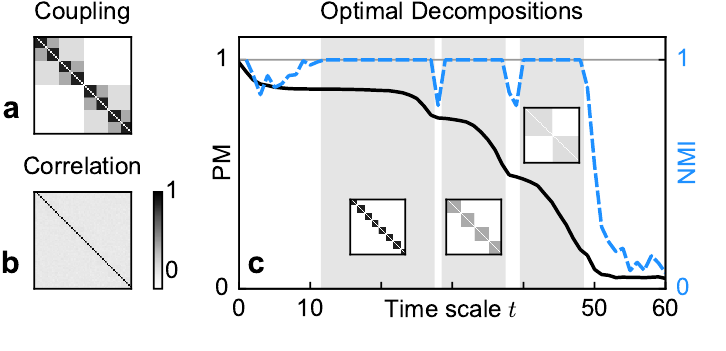}

\caption{(Color online) System of 80 coupled logistic maps ($\alpha{=}2,\gamma{=}0.04$)
\textbf{(a)} The coupling matrix exhibits hierarchical modularity
at three levels. \textbf{(b)} The system is chaotic and no strong
correlations between variables are observed over 10,000 time steps.
\textbf{(c)} Perturbation modularity of optimal decompositions (PM,
solid black) at different time scales $t$ and NMI between optimal
decompositions at successive times (dashed blue). Three time scale
regions are robust (NMI=1, grey), corresponding to the three hierarchical
levels of the coupling matrix (insets).\label{fig:pertmod-fig1}}
\end{figure}

\begin{example}\label{pertmod-example1}

In \textbf{Example \ref{pertmod-example1}}, we uncover modular organization
that is present in a system's coupling matrix, though not apparent
in the correlation statistics. Consider an $N{=}80$ variable system
with chaotic dynamics ($\alpha{=}2,\gamma{=}0.04$) and a hierarchical
modular coupling matrix (Fig.~\ref{fig:pertmod-fig1}a). The system
is composed of 8 tightly-coupled low-level modules ($k_{ji}{=}1)$
with 10 variables each, pairs of which are nested within 4 mid-level
modules ($k_{ji}{=}0.01)$, pairs of which are in turn nested within
2 weakly coupled high-level modules ($k_{ji}{=}0.0001)$. A random
state is used as the initial condition.

Because the system is strongly chaotic for these values of $\alpha$
and $\gamma$, there is no obvious `order parameter' for identifying
modular organization from system trajectories \cite{arenas2006sync};
for instance, variable states are largely uncorrelated over 10,000
time steps (Fig.~\ref{fig:pertmod-fig1}b). However, because perturbations
first spread within low-level modules, then mid-level modules, and
finally high-level modules, our method easily uncovers the hierarchical
modular organization. Fig.~\ref{fig:pertmod-fig1}c shows the perturbation
modularity (black) and NMI (dashed blue) for optimal decompositions
at different time scales. There are three robust time scale regions,
corresponding to each of the three hierarchical levels of the coupling
matrix (insets in Fig.~\ref{fig:pertmod-fig1}c). Beyond time scale
${\sim}50$, perturbations have spread between the high-level modules;
at this point, optimal decompositions reflect random fluctuations
in initial conditions, and perturbation modularity and NMI values
are near 0.

\end{example}

\begin{figure}
\includegraphics[width=1\linewidth]{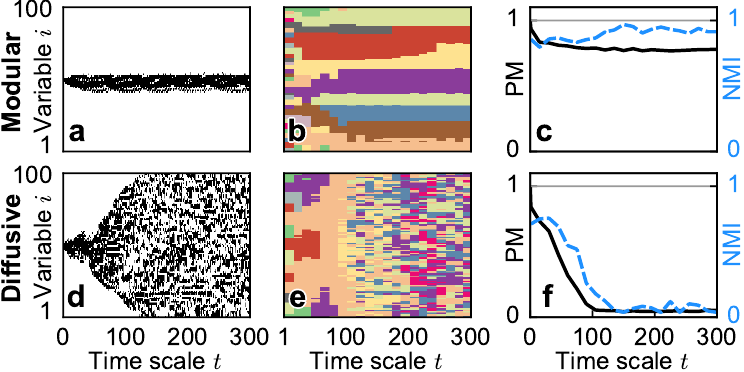}

\caption{(Color online) Two 100-variable CMLs are compared: one `modular' (top
row; $\alpha{=}1.7,\gamma{=}0.1$) and one `diffusive' (bottom row;
$\alpha{=}1.9,\gamma{=}0.6$). \textbf{(a,d)} Spacetime plots of the
effect of a single-variable perturbation. A pixel is colored black
if the absolute difference between perturbed and unperturbed trajectory
at a variable (vertical axis) exceeds 1\% of the size of the system-wide
perturbation at a given time (horizontal axis). \textbf{(b,e)} Spacetime
plots of the optimal decompositions at different time scales; color
indicates each variable's subsystem.\textbf{ (c,f)} Perturbation modularity
of optimal decompositions (PM, solid black) at different time scales
$t$ and NMI between optimal decompositions at successive times (dashed
blue). Stable decompositions are observed in the modular CML (top
row).\label{fig:pertmod-fig2}}
\end{figure}

\begin{example}\label{pertmod-example2}

In \textbf{Example \ref{pertmod-example2}}, we investigate a more
interesting case in which modularity emerges in a homogeneously-coupled
CML. In some parameter regimes, spatial variation in initial conditions
breaks the lattice coupling symmetry and leads to the emergence of
modular \emph{domains} (contiguous lattice regions) that constrain
the spread of perturbations \cite{kaneko1986lyapunov}. Such a `modular'
regime is investigated using a CML with $N{=}100$ variables and weak
coupling-strength ($\alpha{=}1.7,\gamma{=}0.1$). The initial condition
is set by iterating a random state for 10,000 time steps. Fig.~\ref{fig:pertmod-fig2}a
shows the spacetime plot of the effect of a single-variable perturbation
to this initial condition: the perturbation spreads to several nearby
variables until $t{\sim}50$ but then remains confined within its
domain. When computed over a uniform distribution of single-variable
perturbations, our method uncovers robust modular organization for
a large range of time scales (Fig.~\ref{fig:pertmod-fig2}b), with
optimal decompositions exhibiting high values of perturbation modularity
and NMI (Fig.~\ref{fig:pertmod-fig2}c).

The above system can be compared to a CML in a `diffusive' regime
($\alpha{=}1.9,\gamma{=}0.6$). For these parameters, the effects
of perturbations spread freely across the lattice, as shown in the
spacetime plot of Fig.~\ref{fig:pertmod-fig2}d (initial condition
is the same random state as in the modular CML iterated for 10,000
time steps). This system does not exhibit robust modular organization:
optimal decompositions are not stable (Fig.~\ref{fig:pertmod-fig2}e)
and optimal perturbation modularity and NMI values are low (Fig.~\ref{fig:pertmod-fig2}f).
Once the effects of perturbations spread completely around the ring
lattice at $t{\sim}100$, both optimal perturbation modularity and
NMI values are near $0$.

\end{example}

\begin{figure}
\includegraphics[width=1\columnwidth]{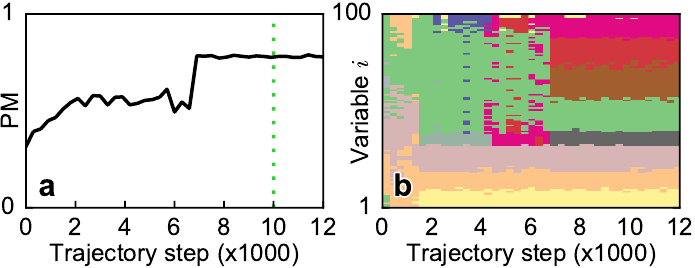}

\caption{(Color online) State-dependence in the modular CML. For a $12,000$
step trajectory starting from a random state, optimal decompositions
(time scale $t{=}300$) are identified using states along this trajectory
as initial conditions. \textbf{(a)} Optimal perturbation modularity
(PM, solid black) grows with increasing trajectory steps (horizontal
axis), indicating the emergence of robust modular structures. Trajectory
step $10,000$ (dotted green line) is the initial condition in examples
\ref{pertmod-example2} and \ref{pertmod-example4}. \textbf{(b)}
Optimal decompositions identified at different trajectory steps; color
indicates the subsystem of each variable (vertical axis).\label{fig:pertmod-transientmod}}
\end{figure}

\begin{example}\label{pertmod-example3}

In \textbf{Example \ref{pertmod-example3}}, we demonstrate state-dependent
modularity by tracking the gradual emergence of modular organization
over the course of a CML trajectory. The modular CML of Example \ref{pertmod-example2}
($\alpha{=}1.7,\gamma{=}0.1$) was started from a random state and
iterated over a $12,000$ step trajectory. The state encountered after
$10,000$ time steps was previously used as the initial condition
in Example \ref{pertmod-example2}. Here we find optimal decompositions
(time scale $t{=}300$) when different states along the aforementioned
trajectory are used as initial conditions. Over the course of the
trajectory, optimal perturbation modularity grows through a series
of plateaus (Fig.~\ref{fig:pertmod-transientmod}a), indicating the
appearance of modular structures. Fig.~\ref{fig:pertmod-transientmod}b
shows the optimal decompositions identified at different trajectory
steps. Variables ${\sim}1{-}40$ quickly form modular structures (by
trajectory step ${\sim}2000$), while variables ${\sim}40{-}100$
need almost $7,000$ steps to do so. This provides an example of \emph{self-organized
modularity}, or modular organization that emerges during a system's
dynamical evolution.

\end{example}

\begin{figure}
\includegraphics[bb=0bp 5bp 210bp 295bp,clip,width=1\columnwidth]{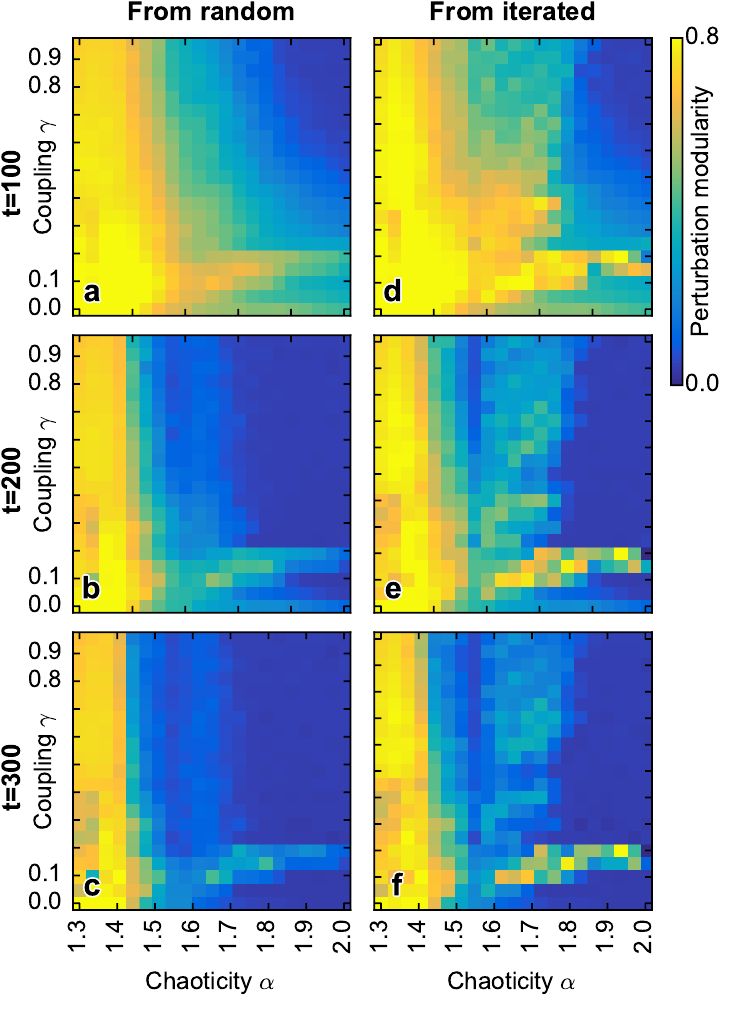}

\caption{(Color online) Parameter phase map of CML. Perturbation modularity
(color) for optimal decompositions of 100-variable CMLs with different
values of chaoticity ($\alpha$; horizontal axes) and coupling strength
($\gamma$; vertical axes). Perturbation modularity is computed at
three time scales for two different classes of initial conditions:
random states\textbf{ {[}(a) }$t{=}100$, \textbf{(b) }$t{=}200$,
and \textbf{(c) }$t{=}300${]} as well as random states iterated for
10,000 time steps \textbf{{[}(d) }$t{=}100$, \textbf{(e) }$t{=}200$,
and \textbf{(f) }$t{=}300${]}.\label{fig:pertmod-phasemap}}
\end{figure}

\begin{example}\label{pertmod-example-phasemap}

Previously, we showed that perturbation modularity captures the presence
of stable modular structures in different CML parameter regimes (Example
\ref{pertmod-example2}), and that it can uncover modular organization
in a state-dependent manner (Example \ref{pertmod-example3}). In
\textbf{Example }\ref{pertmod-example-phasemap}, we use perturbation
modularity to characterize regions in the CML parameter space with
respect to the onset of modularity.

Specifically, we construct 100-variable CMLs with different values
of chaoticity ($\alpha$) and coupling ($\gamma$) parameters. For
these different parameter values, Fig. \ref{fig:pertmod-phasemap}
shows optimal perturbation modularity computed at three time scales
($t{=}100$, $t{=}200$, and $t{=}300$) and two different classes
of initial conditions: random states (Fig. \ref{fig:pertmod-phasemap}a-c)
and random states iterated for 10,000 time steps (Fig. \ref{fig:pertmod-phasemap}d-f).
In all cases, optimal perturbation modularity values were averaged
across 10 random samples of initial conditions. 

Several regimes of spatial organization can be identified in the parameter
phase maps. For $\alpha\lesssim1.44$, spatial domains, which form
even when the system is started from random initial conditions, constrain
the spread of perturbations over long time scales; we call this the
\emph{modular }regime. For other parameter values (e.g., $1.6\lesssim\alpha\lesssim1.95$,
$\gamma\sim0.1$, the yellow `tongue' in Fig. \ref{fig:pertmod-phasemap}d-f),
modular domains only appear when random states are iterated for many
steps before being used as initial conditions. This regime, which
includes the case studied in Example \ref{pertmod-example3}, we call
\emph{self-organized modular}. Finally, for parameter values corresponding
to the blue regions in Fig. \ref{fig:pertmod-phasemap}, which we
call the \emph{diffuse }regime, modular domains are not present and
perturbations spread freely. Here, different parameter values give
rise to different diffusion speeds \cite{kaneko1986lyapunov}: for
example, $\alpha{=}1.9,\gamma{=}0.7$ exhibits no modular organization
at time scale $t{=}100$; on the other hand, $\alpha{=}1.9,\gamma{=}0.2$
maintains some modularity at $t{=}100$, but this organization disintegrates
by $t{=}200$. 

\end{example}

\begin{figure}
\includegraphics[width=1\columnwidth]{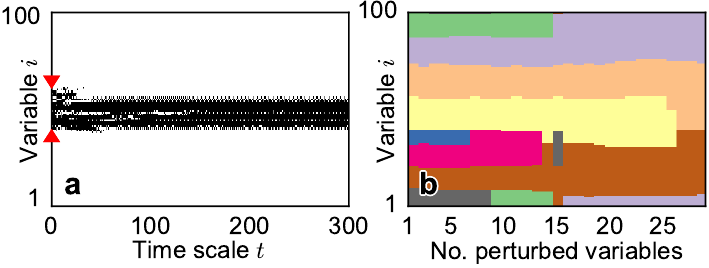}

\caption{(Color online) Perturbation dependence in the modular CML. \textbf{(a)}
Spacetime plot of the effect of a perturbation to a 20-variable window
(red arrows), as in Fig.~\ref{fig:pertmod-fig2}a and \ref{fig:pertmod-fig2}d.
\textbf{(b)} Optimal decompositions (time scale $t{=}300$) for different
perturbation sizes (horizontal axis).\label{fig:pertmod-pertsizes}}
\end{figure}

\begin{example}\label{pertmod-example4}

Finally, in \textbf{Example \ref{pertmod-example4}}, we explore modularity's
dependence on the kinds of perturbations applied. We again consider
the modular CML ($\alpha{=}1.7,\gamma{=}0.1$) and initial condition
of Example \ref{pertmod-example2}. Instead of perturbing single variables,
we now simultaneously perturb multiple variables in lattice-contiguous
`windows' of different sizes (variables simultaneously incremented
by 0.0001; all $N$ windows are perturbed with uniform probability);
for illustration, Fig.~\ref{fig:pertmod-pertsizes}a shows the effect
of a perturbation to a window of \emph{$20$} variables. Fig.~\ref{fig:pertmod-pertsizes}b
shows that optimal decompositions (time scale $t{=}300$) depend on
perturbation size. As more variables are perturbed, smaller subsystems
merge into larger subsystems in a hierarchical manner.

\end{example}

Future work can pursue several extensions to our approach. First,
estimating perturbation modularity from real-world datasets is of
great practical interest; this can be investigated by applying the
method to fitted dynamical models (e.g., vector autoregressive or
dynamical causal modeling \cite{friston2003dynamic}) or using nonparametric
approaches. Second, it is possible to explore other measures of decomposition
quality beyond robustness to time scale, including robustness to changes
in initial conditions and kinds of perturbations; alternatively, decomposition
quality may be evaluated by testing the statistical significance of
optimal perturbation modularity against null-model ensembles of nonmodular
dynamical systems \cite{zalesky2012onthe}. Third, it is of interest
to identify possible limitations of our method, such as whether it
is susceptible to the kinds of resolution limits \cite{fortunato2007resolution}
and detectability thresholds \cite{nadakuditi2012graphspectra} encountered
by graph-based community detection methods. Finally, future research
can investigate other measures of perturbation magnitude (e.g., different
norms or divergence functions), kinds of decompositions (e.g., overlapping
subsystems), and cost functions (beyond vector autocovariance). For
example, cost functions that capture the invertibility or sparsity
of coarse-grained dynamics could be used to decompose a system into
a mesoscopic `control diagram', in which each subsystem controls a
small number of others.

To summarize, we identify modular decompositions of multivariate dynamical
systems based on the idea that modules constrain the spread of perturbations.
We propose a quality function, called perturbation modularity, which
can be used to identify optimal coarse grainings that capture the
slow component of perturbation spreading dynamics. The method generalizes
graph-based community detection to a broad class of nonlinear dynamical
systems and provides a principled alternative to detecting communities
in network representations of dynamics. The method captures variation
in modular organization across different time scales, initial conditions,
and kinds of perturbations and offers a powerful tool for exploring
modularity in complex systems.

\medskip{}

\vspace{-10mm}
\begin{acknowledgments}
\vspace{-5mm}
We thank Y-Y Ahn, Filippo Radicchi, and Daniel Damineli for helpful
feedback. This work was supported in part by a NSF IGERT fellowship
to AJG, Fundação para a Ciencia e a Tecnologia (Portugal) grant PTDC/EIA-CCO/114108/2009,
and the FLAD Computational Biology Collaboratorium (Portugal).
\end{acknowledgments}

\bibliographystyle{apsrev4-1}
\bibliography{main}

\newpage

\onecolumngrid
\appendix

\renewcommand{\thesubsection}{\thesection\arabic{subsection}}

\part*{Supplemental Material}

\section{Bounds on Perturbation Modularity}

The bounds on perturbation modularity depend on the norm used to measure
perturbation magnitude (see Eq. (1) in the main text). Here we consider
the family of $\ell_{p}$ norms, where the $\ell_{p}$ norm of vector
$\bm{a}=\left\langle a_{1},\dots,a_{N}\right\rangle $ is defined
for $p>0$ as ${\textstyle \left\Vert \bm{a}\right\Vert _{p}=\left(\sum_{i=1}^{N}\left|a_{i}\right|^{p}\right)^{\nicefrac{1}{p}}}$.

As we show below, perturbation modularity is bounded by $-1$ and
$1$ for $\ell_{1}$ and $\ell_{2}$ norms. More generally, for $\ell_{p}$
norms with $p\ge2$, perturbation modularity is bounded by $-N^{\left(\nicefrac{p-2}{p}\right)}$
and $N^{\left(\nicefrac{p-2}{p}\right)}$, where $N$ is the number
of variables in the original system.

Assume some $\ell_{p}$ norm is used to measure perturbation magnitude.
Using the definition of perturbation modularity for initial condition
$\bm{x}$ and partition $\pi$:
\begin{eqnarray*}
Q^{t}(\pi,\bm{x}) & = & \textrm{E}\!\left[\bm{y}_{\pi}^{0}(\bm{x},\bm{\varepsilon})\cdot\bm{y}_{\pi}^{t}(\bm{x},\bm{\varepsilon})\right]-\mathbb{\textrm{E}}\!\left[\bm{y}_{\pi}^{0}(\bm{x},\bm{\varepsilon})\right]\cdot\textrm{E}\!\left[\bm{y}_{\pi}^{t}(\bm{x},\bm{\varepsilon})\right]\\
 & \le & \textrm{E}\!\left[\bm{y}_{\pi}^{0}(\bm{x},\bm{\varepsilon})\cdot\bm{y}_{\pi}^{t}(\bm{x},\bm{\varepsilon})\right]\\
 & \le & \textrm{E}\!\left[\left\Vert \bm{y}_{\pi}^{0}(\bm{x},\bm{\varepsilon})\right\Vert _{2}\left\Vert \bm{y}_{\pi}^{t}(\bm{x},\bm{\varepsilon})\right\Vert _{2}\right]
\end{eqnarray*}
where the second line follows from the non-negativity of the coarse-grained
perturbation vectors $\bm{y}_{\pi}^{0}(\bm{x},\bm{\varepsilon})$
and $\bm{y}_{\pi}^{t}(\bm{x},\bm{\varepsilon})$ and the third line
follows from the Cauchy\textendash Schwarz inequality. 

When some $\ell_{p}$ norm is used to measure perturbation magnitude,
the coarse-grained perturbation vectors themselves are unit vectors
in $\ell_{p}$. To show this, let $v_{i}=|\left(f^{t}(\bm{x}+\bm{\varepsilon})-f^{t}(\bm{x})\right)_{\{i\}}|$.
Then, 
\begin{eqnarray*}
\left\Vert \bm{y}_{\pi}^{t}(\bm{x},\bm{\varepsilon})\right\Vert _{p} & = & \left(\sum_{S\in\pi}\left[m_{S}^{t}(\bm{x},\bm{\varepsilon})\right]^{p}\right)^{1/p}=\left(\sum_{S\in\pi}\left[\frac{\left(\sum_{i\in S}v_{i}^{p}\right)^{\frac{1}{p}}}{\left(\sum_{i^{\prime}=1}^{N}v_{i^{\prime}}^{p}\right)^{\frac{1}{p}}}\right]^{p}\right)^{1/p}=\left(\sum_{S\in\pi}\frac{\sum_{i\in S}v_{i}^{p}}{\sum_{i^{\prime}=1}^{N}v_{i^{\prime}}^{p}}\right)^{1/p}=1
\end{eqnarray*}

When $p=2$ is used to measure perturbation magnitudes, the upper
bound can be rewritten as: 
\begin{eqnarray*}
Q^{t}\left(\pi,\bm{x}\right)\le\textrm{E}\!\left[\left\Vert \bm{y}_{\pi}^{0}(\bm{x},\bm{\varepsilon})\right\Vert _{2}\left\Vert \bm{y}_{\pi}^{t}(\bm{x},\bm{\varepsilon})\right\Vert _{2}\right]=\textrm{E}\!\left[1\cdot1\right]=1
\end{eqnarray*}

When $p=1$ is used to measure perturbation magnitudes, we first note
that $\left\Vert \bm{a}\right\Vert _{2}\le\left\Vert \bm{a}\right\Vert _{1}$
for any $\bm{a}$ and that $\left\Vert \bm{y}_{\pi}^{0}(\bm{x},\bm{\varepsilon})\right\Vert _{1}=\left\Vert \bm{y}_{\pi}^{t}(\bm{x},\bm{\varepsilon})\right\Vert _{1}=1$.
The upper bound can be rewritten as:
\begin{eqnarray*}
Q^{t}\left(\pi,\bm{x}\right)\le\textrm{E}\!\left[\left\Vert \bm{y}_{\pi}^{0}(\bm{x},\bm{\varepsilon})\right\Vert _{2}\left\Vert \bm{y}_{\pi}^{t}(\bm{x},\bm{\varepsilon})\right\Vert _{2}\right]\le\textrm{E}\!\left[1\cdot1\right]=1
\end{eqnarray*}

More generally, when $p\ge2$, Hölder's inequality \cite{tao2010epsilon}
provides the bound $\left\Vert \bm{a}\right\Vert _{2}\le n^{\left(\frac{1}{2}-\frac{1}{p}\right)}\left\Vert \bm{a}\right\Vert _{p}$,
where $n$ is the number of dimensions of $\bm{a}$. Thus, $\left\Vert \bm{y}_{\pi}^{t}(\bm{x},\bm{\varepsilon})\right\Vert _{2}\le\left|\pi\right|^{\left(\frac{1}{2}-\frac{1}{p}\right)}\left\Vert \bm{y}_{\pi}^{t}(\bm{x},\bm{\varepsilon})\right\Vert _{p}=\left|\pi\right|^{\left(\frac{1}{2}-\frac{1}{p}\right)}\le N^{\left(\frac{1}{2}-\frac{1}{p}\right)}$,
since $N$ is the maximum number of subsets in a partition. This gives:
\begin{eqnarray*}
Q^{t}\left(\pi,\bm{x}\right)\le\textrm{E}\!\left[\left\Vert \bm{y}_{\pi}^{0}(\bm{x},\bm{\varepsilon})\right\Vert _{2}\left\Vert \bm{y}_{\pi}^{t}(\bm{x},\bm{\varepsilon})\right\Vert _{2}\right]\le\textrm{E}\!\left[N^{\left(\frac{1}{2}-\frac{1}{p}\right)}N^{\left(\frac{1}{2}-\frac{1}{p}\right)}\right]=N^{\left(\frac{p-2}{p}\right)}
\end{eqnarray*}

To show the lower bound, we again use the definition of perturbation
modularity and similar reasoning:
\begin{eqnarray*}
Q^{t}(\pi,\bm{x}) & = & \textrm{E}\!\left[\bm{y}_{\pi}^{0}(\bm{x},\bm{\varepsilon})\cdot\bm{y}_{\pi}^{t}(\bm{x},\bm{\varepsilon})\right]-\mathbb{\textrm{E}}\!\left[\bm{y}_{\pi}^{0}(\bm{x},\bm{\varepsilon})\right]\cdot\textrm{E}\!\left[\bm{y}_{\pi}^{t}(\bm{x},\bm{\varepsilon})\right]\\
 & \ge & -\mathbb{\textrm{E}}\!\left[\bm{y}_{\pi}^{0}(\bm{x},\bm{\varepsilon})\right]\cdot\textrm{E}\!\left[\bm{y}_{\pi}^{t}(\bm{x},\bm{\varepsilon})\right]\\
 & \ge & -\left\Vert \mathbb{\textrm{E}}\!\left[\bm{y}_{\pi}^{0}(\bm{x},\bm{\varepsilon})\right]\right\Vert _{2}\left\Vert \textrm{E}\!\left[\bm{y}_{\pi}^{t}(\bm{x},\bm{\varepsilon})\right]\right\Vert _{2}\\
 & \ge & -\textrm{E}\!\left[\left\Vert \bm{y}_{\pi}^{0}(\bm{x},\bm{\varepsilon})\right\Vert _{2}\right]\textrm{E}\!\left[\left\Vert \bm{y}_{\pi}^{0}(\bm{x},\bm{\varepsilon})\right\Vert _{2}\right]
\end{eqnarray*}
where the last line follows from Jensen's inequality and the convexity
of norms. Similar arguments as above show that $Q^{t}\left(\pi,\bm{x}\right)\ge-1$
for $p=1$ and $p=2$ and $Q^{t}\left(\pi,\bm{x}\right)\ge-N^{\left(\nicefrac{p-2}{p}\right)}$
for $p>2$.

In practice, we are also interested in the maximal perturbation modularity
across all partitions. In fact, maximal perturbation modularity is
always non-negative. This is because there is at least one partition
with perturbation modularity equal to $0$: the partition that contains
the entire system as one subsystem. In this partition, which we call
$\pi_{0}=\left\{ \left\{ 1,\dots,N\right\} \right\} $, perturbations
are entirely contained in the single subsystem and $\bm{y}_{\pi_{0}}^{0}(\bm{x},\bm{\varepsilon})=\bm{y}_{\pi_{0}}^{t}(\bm{x},\bm{\varepsilon})=\left\langle 1\right\rangle $
for all $\bm{x}$ and $\bm{\varepsilon}$. From the definition of
perturbation modularity, it can be seen that $Q^{t}\left(\pi_{0},\bm{x}\right)=0$
for all $t$.

\newpage{}

\section{Perturbation modularity and community detection}

An explicit connection can be made between our approach and graph-based
community detection. We first rewrite and expand the definition of
perturbation modularity from the main text:
\begin{eqnarray*}
Q^{t}(\pi,\bm{x}) & = & \textrm{E}\!\left[\bm{y}_{\pi}^{0}(\bm{x},\bm{\varepsilon})\cdot\bm{y}_{\pi}^{t}(\bm{x},\bm{\varepsilon})\right]-\mathbb{\textrm{E}}\!\left[\bm{y}_{\pi}^{0}(\bm{x},\bm{\varepsilon})\right]\cdot\textrm{E}\!\left[\bm{y}_{\pi}^{t}(\bm{x},\bm{\varepsilon})\right]\\
 & = & \sum_{S\in\pi}\left(\textrm{E}\!\left[m_{S}^{0}(\bm{x},\bm{\varepsilon})\thinspace m_{S}^{t}(\bm{x},\bm{\varepsilon})\right]-\textrm{E}\!\left[m_{S}^{0}(\bm{x},\bm{\varepsilon})\right]\textrm{E}\left[m_{S}^{t}(\bm{x},\bm{\varepsilon})\right]\right)
\end{eqnarray*}
where the expectations are over $P(\bm{\varepsilon})$. We assume
that the $\ell_{1}$ norm is used to measure perturbation magnitude,
which provides the following additive property: $m_{S}^{t}(\bm{x},\bm{\varepsilon})=\sum_{i\in S}m_{\{i\}}^{t}(\bm{x},\bm{\varepsilon})$,
where the subscript $\left\{ i\right\} $ indicates a subsystem only
containing variable $i$. We rewrite the above equation as:
\begin{eqnarray}
Q^{t}(\pi,\bm{x}) & = & \sum_{S\in\pi}\left(\textrm{E}\!\left[\sum_{i\in S}m_{\{i\}}^{0}(\bm{x},\bm{\varepsilon})\sum_{j\in S}m_{\{j\}}^{t}(\bm{x},\bm{\varepsilon})\right]-\textrm{E}\!\left[\sum_{i\in S}m_{\{i\}}^{0}(\bm{x},\bm{\varepsilon})\right]\textrm{E}\!\left[\sum_{j\in S}m_{\{j\}}^{t}(\bm{x},\bm{\varepsilon})\right]\right)\nonumber \\
 & = & \sum_{S\in\pi}\left(\sum_{i,j\in S}\textrm{E}\!\left[m_{\{i\}}^{0}(\bm{x},\bm{\varepsilon})\thinspace m_{\{j\}}^{t}(\bm{x},\bm{\varepsilon})\right]-\sum_{i\in S}\textrm{E}\!\left[m_{\{i\}}^{0}(\bm{x},\bm{\varepsilon})\right]\sum_{j\in S}\textrm{E}\!\left[m_{\{j\}}^{t}(\bm{x},\bm{\varepsilon})\right]\right)\label{eq:pertmod-supp-q2}
\end{eqnarray}
where the last line comes from exchanging the order of summation and
expectation.

\subsection{Perturbation modularity is Markov Stability for diffusion dynamics }

\emph{Markov stability} is a method of community detection that uses
the dynamics random walks over graphs \cite{delvenne2010stability,lambiotte2014random}.
Here, communities are defined as subgraphs that tend to trap random
walkers. This method is of particular interest because it generalizes
many other community detection methods, including optimization of
Newman's modularity, cut size, and spectral methods \cite{lambiotte2014random}.
Given a random walk over an $N$-node graph, the Markov stability
of a partition $\pi$ at time scale $t$ is defined as:
\begin{eqnarray}
R^{t}(\pi) & = & \sum_{S\in\pi}\Pr(\textrm{walker in \emph{S} at times \emph{0} and \emph{t}})-\Pr(\textrm{walker in \emph{S} at time 0})\Pr(\textrm{walker in \emph{S} at time \ensuremath{t}})\label{eq:pertmod-suppmat-markovstability}
\end{eqnarray}
In this framework, the optimal partition of a graph is the one that
maximizes Markov stability.

Given homogenous perturbation to single variables, there is an equivalence
between Markov stability and $\ell_{1}$ perturbation modularity of
diffusion dynamics. Specifically, the diffusion of random walkers
on a graph can be stated in terms of a linear dynamical system $f^{t}\left(\bm{x}\right)=T^{t}\bm{x}$,
where $\bm{x}\left(t\right)$ is an $N$-dimensional vector with $\bm{x}_{i}\left(t\right)$
being the density of random walkers at node $i$ at time $t$, and
$T^{t}$ is the time scale $t$ transition matrix (here superscript
$t$ indicates matrix power; for simplicity, we consider the discrete-time
case). Assume that perturbation to variable $i$ is indicated by $\bm{\varepsilon}^{i}=\mathbf{e}_{i}$,
where $\mathbf{e}_{i}$ is the $i^{\mathrm{th}}$ $N$-dimensional
standard basis vector (initial perturbations can be scaled by any
constant without changing results). Then, $m_{\{j\}}^{0}(\bm{x},\bm{\varepsilon}^{i})=\frac{\left|\bm{\varepsilon}_{\{j\}}^{i}\right|}{\left\Vert \bm{\varepsilon}^{i}\right\Vert _{1}}=\frac{\left|(\mathbf{e}_{i})_{\{j\}}\right|}{\left\Vert \mathbf{e}_{i}\right\Vert _{1}}=\delta_{i,j}$. 

$T^{t}$ is a transition matrix: it has positive entries and preserves
the $\ell_{1}$ norm upon matrix multiplication. Therefore:
\begin{eqnarray*}
m_{\{j\}}^{t}(\bm{x},\bm{\varepsilon}^{i}) & = & \frac{\left|\left(f^{t}(\bm{x}+\bm{\varepsilon}^{i})-f^{t}(\bm{x})\right)_{\{j\}}\right|}{\left\Vert f^{t}(\bm{x}+\bm{\varepsilon}^{i})-f^{t}(\bm{x})\right\Vert _{1}}=\frac{\left|(T^{t}(\bm{x}+\bm{\varepsilon}^{i})-T^{t}\bm{x})_{\{j\}}\right|}{\left\Vert T^{t}(\bm{x}+\bm{\varepsilon}^{i})-T^{t}\bm{x}\right\Vert _{1}}=\frac{\left|(T^{t}\bm{\varepsilon}^{i})_{\{j\}}\right|}{\left\Vert T^{t}\bm{\varepsilon}^{i}\right\Vert _{1}}=\frac{\left|(T^{t}\mathbf{e}_{i})_{\{j\}}\right|}{\left\Vert T^{t}\mathbf{e}_{i}\right\Vert _{1}}=T_{ij}^{t}
\end{eqnarray*}

The terms in Eq.~(\ref{eq:pertmod-supp-q2}) can now be mapped to
the terms in the Markov stability Eq.~(\ref{eq:pertmod-suppmat-markovstability}):
\begin{eqnarray*}
\sum_{i,j\in S}\textrm{E}\!\left[m_{\{i\}}^{0}(\bm{x},\bm{\varepsilon})\thinspace m_{\{j\}}^{t}(\bm{x},\bm{\varepsilon})\right] & = & \sum_{i,j\in S}\sum_{k=1}^{N}P(\bm{\varepsilon}^{k})\delta_{i,k}T_{kj}^{t}=\sum_{i,j\in S}P(\bm{\varepsilon}^{i})T_{ij}^{t}=\Pr(\textrm{walker in \emph{S} at times \emph{0} and \emph{t}})\\
\sum_{i\in S}\textrm{E}\!\left[m_{\{i\}}^{0}(\bm{x},\bm{\varepsilon})\right] & = & \sum_{i\in S}\sum_{k=1}^{N}P(\bm{\varepsilon}^{k})\delta_{i,k}=\sum_{i\in S}P(\bm{\varepsilon}^{i})=\Pr(\textrm{walker in \emph{S} at time \emph{0}})\\
\sum_{j\in S}\textrm{E}\!\left[m_{\{j\}}^{t}(\bm{x},\bm{\varepsilon})\right] & = & \sum_{j\in S}\sum_{k=1}^{N}P(\bm{\varepsilon}^{k})T_{kj}^{t}=\Pr(\textrm{walker in \emph{S} at time \emph{t}})
\end{eqnarray*}

Markov stability is usually defined for an equilibrium random walk,
such that $p\left(\textrm{walker in \emph{S} at time 0}\right)=p\left(\textrm{walker in \emph{S} at time \emph{t}}\right)$.
In our framework, this is accomplished by setting $P\left(\bm{\varepsilon}^{i}\right)$
equal to the equilibrium probability of finding a random walker at
node $i$.

\subsection{Mapping to directed weighted Newman's modularity}

In this section, we show that for any dynamical system, $\ell_{1}$
perturbation modularity can be mapped to \emph{directed weighted Newman's
modularity} on a specially-constructed graph. One result of this mapping
is that efficient community detection algorithms can be used to find
partitions that maximize perturbation modularity.

To review, the weighted directed Newman's modularity of partition
$\pi$ is defined as \cite{arenas2007sizereduction}:
\begin{equation}
\sum_{C\in\pi}\sum_{i,j\in C}\left(w_{ij}-\frac{w_{i}^{out}w_{j}^{in}}{M}\right)\label{eq:directmodularity}
\end{equation}
where $w_{ij}$ indicates edge weight from node $i$ to node $j$,
$w_{i}^{out}=\sum_{j}w_{ij}$ is the out-degree, $w_{j}^{in}=\sum_{i}w_{ij}$
is the in-degree, and $M=\sum_{i}\sum_{j}w_{ij}=\sum_{i}w_{i}^{out}=\sum_{j}w_{j}^{in}$
is the total strength (summations in these equations are over all
nodes).

Now, for an $N$-dimensional dynamical system, we construct a graph
with one node for each variable and edge weight from node $i$ to
$j$:
\begin{eqnarray*}
w_{ij} & = & \textrm{E}\!\left[m_{\{i\}}^{0}(\bm{x},\bm{\varepsilon})\thinspace m_{\{j\}}^{t}(\bm{x},\bm{\varepsilon})\right]
\end{eqnarray*}
When the $\ell_{1}$ norm is used, $\sum_{i=1}^{N}m_{\{i\}}^{t}(\bm{x},\bm{\varepsilon})=1$.
Thus, 
\begin{eqnarray*}
w_{i}^{out} & = & \sum_{j=1}^{N}w_{ij}=\sum_{j=1}^{N}\textrm{E}\!\left[m_{\{i\}}^{0}(\bm{x},\bm{\varepsilon})\thinspace m_{\{j\}}^{t}(\bm{x},\bm{\varepsilon})\right]=\textrm{E}\!\left[m_{\{i\}}^{0}(\bm{x},\bm{\varepsilon})\sum_{j=1}^{N}m_{\{j\}}^{t}(\bm{x},\bm{\varepsilon})\right]=\textrm{E}\!\left[m_{\{i\}}^{0}(\bm{x},\bm{\varepsilon})\right]\\
w_{j}^{in} & = & \sum_{i=1}^{N}w_{ij}=\sum_{i=1}^{N}\textrm{E}\!\left[m_{\{i\}}^{0}(\bm{x},\bm{\varepsilon})\thinspace m_{\{j\}}^{t}(\bm{x},\bm{\varepsilon})\right]=\textrm{E}\!\left[\left(\sum_{i=1}^{N}m_{\{i\}}^{0}(\bm{x},\bm{\varepsilon})\right)m_{\{j\}}^{t}(\bm{x},\bm{\varepsilon})\right]=\textrm{E}\!\left[m_{\{j\}}^{t}(\bm{x},\bm{\varepsilon})\right]\\
M & = & \sum_{i=1}^{N}w_{i}^{out}=\sum_{i=1}^{N}\textrm{E}\!\left[m_{\{i\}}^{0}(\bm{x},\bm{\varepsilon})\right]=\textrm{E}\!\left[\sum_{i=1}^{N}m_{\{i\}}^{0}(\bm{x},\bm{\varepsilon})\right]=\textrm{E}\!\left[1\right]=1
\end{eqnarray*}

Rewriting Eq.~(\ref{eq:pertmod-supp-q2}) makes the mapping to Eq.~(\ref{eq:directmodularity})
explicit: 
\begin{eqnarray*}
Q^{t}(\pi,\bm{x}) & = & \sum_{S\in\pi}\left(\sum_{i,j\in S}\textrm{E}\!\left[m_{\{i\}}^{0}(\bm{x},\bm{\varepsilon})\thinspace m_{\{j\}}^{t}(\bm{x},\bm{\varepsilon})\right]-\sum_{i\in S}\textrm{E}\!\left[m_{\{i\}}^{0}(\bm{x},\bm{\varepsilon})\right]\sum_{j\in S}\textrm{E}\!\left[m_{\{j\}}^{t}(\bm{x},\bm{\varepsilon})\right]\right)\\
 & = & \sum_{S\in\pi}\sum_{i,j\in S}\left(\textrm{E}\!\left[m_{\{i\}}^{0}(\bm{x},\bm{\varepsilon})\thinspace m_{\{j\}}^{t}(\bm{x},\bm{\varepsilon})\right]-\textrm{E}\!\left[m_{\{i\}}^{0}(\bm{x},\bm{\varepsilon})\right]\textrm{E}\!\left[m_{\{j\}}^{t}(\bm{x},\bm{\varepsilon})\right]\right)\\
 & = & \sum_{S\in\pi}\left(\sum_{i,j\in S}w_{ij}-\frac{w_{i}^{out}w_{j}^{in}}{M}\right)
\end{eqnarray*}


\end{document}